# Cheat proof Communication through Cluster Head (C3H) in Mobile Ad Hoc Network

Abu Sufian, Anuradha Banerjee and Paramartha Dutta

Abu Sufian is with University of Gour Banga, West Bengal, India (e-mail: sufian.csa@gmail.com)

Anuradha Banerjee is with the Kalyani Govt. Engg. College, West Bengal, India (e-mail: anuradha79bn@gmail.com).

Paramartha Dutta is Visva-Bharati University, West Bengal, India(e-mail: paramartha.dutta@gmail.com)

*Abstract:*

*Mobile ad hoc network (MANET) is a wireless network based on a group of mobile nodes without any centralized infrastructure. In civilian data communication all nodes cannot be homogeneous type and not do a specific data communication. Therefore, node co-operation and cheat proof are essential to successfully run MANETs in civilian data communications. Denial of service and malicious behavior of the node are the main concern to secure and successful communication in MANETs. This scheme proposed a generic solution to prevent malicious behavior of the node by the cluster head through single hop node clustering strategy.*

## 1. Introduction

Mobile ad-hoc networks (MANETs) is an infrastructure less, self-organizing network where a set of mobile nodes (capable of receiving and transmitting radio signals) can quickly set up a temporary network [1]. This type of networks is very useful in emergency situation like a battle field, rescue operation after natural disaster, commercial application like vehicular ad hoc networks, communication in conference hall and many more [2]. There are many underlined protocols are available to establish this type of network [3]. We can classify all these protocols into three broad categories: one is proactive or table driven such as DSDV [4], WRP [5], FSR [6], etc., second type is reactive such as AODV [7], AOMDV [8], DSR [9], etc. and the third category is hybrid where some part of the network is proactive and other part reactive, such as EMR-PL [10], TORA [11], etc.

In most of the routing protocol network security and lack of co-operation among nodes are yet to be solved. It is true fact that MANETs is not successful in civilian data communication because of lack of co-operations and malicious behavior of some nodes, although it is successful in specific communication such as military data communication. Attacks by some malicious nodes from inside of the network are a main security issue in MANETs. This malicious behavior is not an issue of military communication or any other specific communication of MANETs, because all nodes of such type of communication are same type and they are specifically working in that communication. But in general life data communication and networking, mobile nodes are open and types are different such as cell phone, laptop, palmtop, PDA etc. Therefore, network security is very much important here.

In our scheme of cheat proof communication through cluster head (C3H), we use single hop clustering strategies as a generic mechanism to increase network security through cluster head (CH), so that MANETs could be successful in civilian data communication. We know that clustering scheme is more scalable and we can see it in FESC [12]. In single hop clustering all the nodes attached with CH directly and makes the entire network into different partitioned called clusters. CH are connected to each other through gateway nodes and established a MANETs and this CHs will take responsibility of network security and increase co-operation within networks. Network security is a big challenge from starting time of MANETs [13]. All the network security and co-operation attacks in MANETs can be classified into two broad categories: one is selfish attack and another one is a malicious attack. Malicious attack is more harmful than selfish attack in this type of networks. This scheme explains the idea to make prevention, such malicious attacks one by one through CHs.

The rest of the paper organized as follows: In section 2 similar works has studied, section-3 explain clustering strategy, our proposed solution to malicious attacks has shown in section-4, in section-5 discussion of simulation results and conclusion has drawn in section-7.

## 2. Literature survey

Many node co-operation and cheat proof scheme proposed by many researchers so far. Some prominent schemes are as follows:

In [14] Thomas J. Giuli and Mary G. Baker suggest a routing scheme based on DSR [9] to detect misbehaving node using 'watchdog' and give labels using 'pathrater'. By this 'pathrater' nodes are classified, and then misbehaving or malicious node avoided. Levente Buttyan and Jean-Pierre Hubaux proposed a virtual currency based scheme called 'Nuglet'[15] to increase node co-operation in MANETs. The researchers of this scheme used two purse models; one is a Packet Purse Model (PPM) where 'Nuglet' debited from the source of the packet, and another is a Packet Trade Model (PTM) where 'Nuglet' debited from the source of the packet, and another is a Packet Trade Model (PTM) where 'Nuglet' debited from the destination of the packet. This scheme also described the purpose of increasing 'Nuglet' for a node, and also discussed security of these 'Nuglet'. Pietro Michiardi and Refik Molva proposed a reputation based scheme called CORE [16]. This scheme stimulates selfish node to avoid selfish behavior such as denial of service attack. In [17] Sheng Zhong et. al proposed a simple, cheat-proof, credit-based system for mobile ad-hoc networks with selfish nodes called SPRITE. This is an incentive credit or debit based system without any tamper proof hardware. Here node can get inceptive by showing receipt of forwarded message from Credit Clearance Service (CCS). Frank Kargl et.al proposed Advanced Detection of Selfish or Malicious Nodes in Ad hoc Networks [18]. This scheme explains activity-based overhearing, iterative probing and unambiguous probing to detect malicious and selfish nodes in the network. In [19] N. Nasser and Y. Chen proposed an intrusion detection scheme. Here malicious nodes detected by overhearing the network then give responses. This is an enhanced version of 'watchdog' and 'pathrater'. In [20] Nan Kang et.al present a misbehaving node detection scheme at IIWAS2010. They used different Intrusion Detection System (IDS) alternate to watchdog to detect malicious node. This IDS is called Enhanced Adaptive ACKnowledgement (EAACK) which had tried to overcome difficulties of watchdog.

Enrique Hernandez-Orallo et.al proposed cocoa as a collaborative contact based 'watchdog'[21] to effectively detect selfish nodes with less time. This scheme said to depend on 'watchdog' only is too much expectation, rather CoCoWa use collaborative work, based on the diffusion of local selfish nodes awareness. Jian-Ming Chang et.al proposed a Cooperative Bait Detection Approach CBDS [22] based on DSR [9]. This scheme exploits both proactive and reactive defense architectures. Here authors used reverse tracing approach to defend a collaborative attack by malicious nodes. In [23] Sara Berri et.al present reputation based node cooperation at an international conference. According to their scheme co-operation of node can be increased by adding or deducting reputation of that node within the network. If a node denied giving service to reputed node, then the most reputation will lose compare to deny servicing less reputed node. Similarly could gain more reputation by serving reputed node and less for non-reputed node.

## 3. Clustering scheme details

The main problem of MANETs is the mobility of node and for that reason topology is very dynamic. Due to this high mobility traditional protocols and security scheme of fixed network is not working in ad hoc networks. As a clustering strategy could mimic the topology of traditional network and it reduces the scalability problem which is very essential for MANETs and it can also reduce other problematic issues in MANETs. Here we have adopted FESC [12] that is a single hop clustering scheme with some modifications for cheat proof and co-operation among nodes in MANETs. There are three types of nodes in this scheme namely Cluster Head(CH), gateway node and ordinary member. Single hop clustering scheme is a strategy where all the mobile nodes attached with some elected CHs directly making the entire network of many groups of nodes headed by each CH. CHs are elected temporarily according to high residual energy, bandwidth and low mobility of node compare to other node of a cluster. A portion of an instance of our clustering scheme has been shown in figure-1.

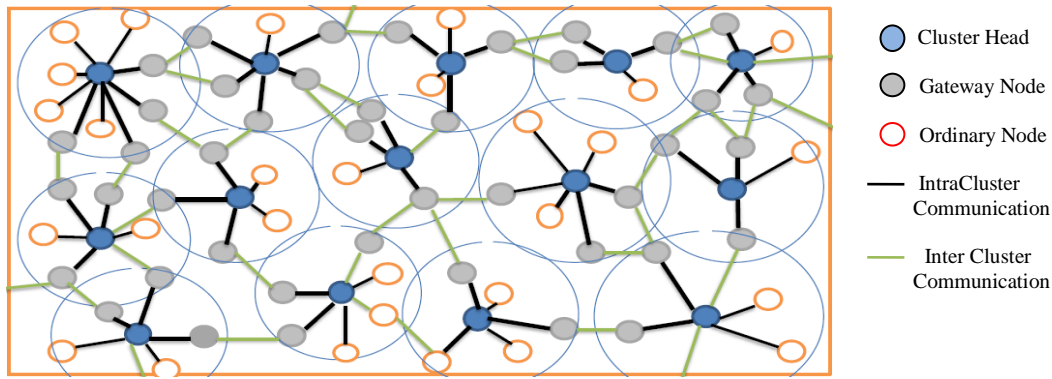

Figure-1: an instance of our clustering

### 3.1 Cluster formation strategy

The most important strategy in clusters scheme is the election of Cluster Head (CH). The stability of the routing directly depends on stability of CHs. In this scheme, we assume CHs are fully supportive and trusted nodes, therefore the importance of choosing a good candidate for CH is very high. This is a single hop clustering scheme and other important strategy consisting adding a node to a cluster, deleting a node from the cluster and merging two clusters to form a new cluster same as FESC [12].

### *3.2 Electing Custer Head (CH)*

Four important factors of node are measured and combine with the help of fuzzy logic and get final metric. This final metric is used to elect CH and gateway node. The four important factors or metrics are: residual energy, trust value , mobility of a node and connectivity to downlink neighbors.

**3.2.1 Residual energy of node:** According to functionality of node at least 40% residual energy required to remain operational. Let $E_i$ is the total residual energy of node $n_i$, $e_i$ is the expanded energy till current time and rem_eng(i) is the current residual energy as a fuzzy variable with values between 0 and 1. Therefore, the current residual energy represented by the equation (1).

$$res\_eng(i) = 1 - \frac{e_i}{E_i} \qquad (1)$$

The value of '*res_eng*' below 0.4 means worst and near 1 means best.

**3.2.2 Trust Value of Node:** At the initial stage that is when a node enters the network, 0.5 assign as a default trust value, where '*trust_value*' is a fuzzy variable, values 0 to 1 indicating the trusted level of a node. There are two more supporting variables, one is '*earn_trust*' which can take any natural number starting from 2 and another is '*loose_trust*' which is also a natural number ranges from 1 up to '*earn_trust*'. Initial values assigned to '*earn_trust*' and '*loose_trust*' are 2 and 1 respectively. When a node successfully transfers a packet its '*earn_trust*' value increases by 1 if any kind of selfish behavior shown its '*loose_trust*' value decreases by 1 and if any, malicious behavior shown, then it lose all '*trust_value*', so that its overall trust values decreases to 0. Trust value will also decrease after successful completion of the data transfer request by one unit. A node can ask to CH to transfer its data packet only if it has positive trust value. The entire activity is carried out by CH. The equation (2) is used to calculate current '*trust_value*' of a node.

$$trust\_value = 1 - \frac{loose\_trust}{earn\_trust} \qquad (2)$$

More trust means more chances to become CH as well as more credit to send data packet.

**3.2.3 Mobility of node:** In order to get the stable clustering scheme in MANETs, the CH of each cluster should less mobile with its downlink neighbors compare to other nodes of the same cluster. Let transmission power of a signal of node $n_a$ is *trans_power(a)* and power of this signal when it being received at node $n_i$ is *recv_power$_b$(a)* and current distance between node $n_a$ and $n_b$ at the time of i-th Hello message is *dist$_i$(a,b)*. Therefore, as per Frii's transmission equation:

$$recv\_power_b(a) = K. trans\_power(a)/ dist^q(a,b) \qquad (3)$$

where K is constant and q is a factors with values 2,3 or 4 depending upon environment. Re-write the equation (3) and get

$$dist_i(a,b) = \sqrt[q]{K. trans\_power(a) / recv\_power(a)} \qquad (4)$$

Suppose t be the time interval between two consecutive HELLO message and n is the number of HELLO message observed. Therefore, the effective mobility of node $n_a$ compared to its down link neighbors is calculated by equation (5). The average mobility called '*avg_mobility*' of a node $n_a$ with respect to all its downlink neighbors is in the equation (6).

$$mobility_b(a) = \{\sum_{i=2}^{n} (dist_i(a,b) - dist_{i-1}(a,b))\} / (n \times t) \quad (5)$$

$$avg\_mobility(a) = \sum_{K=1}^{n_d} mobility(a) / k \quad (6)$$

**3.2.4 Downlink neighbors connectivity:** The CH should have more downlink neighbors compare to other member node. Here we assume that the current CH has the standard number of downlink neighbors. This number of downlink neighbors of CH calculated by the fuzzy membership value of that CH, and initial standard membership value of this parameter are 0.5. Therefore, any node which has more number of downlink neighbors has more chances to become CH.

Let number of downlink neighbors of current CH is '*ndnb_CH*', number of downlink neighbors of node $n_i$ are '*ndnb_$n_i$*'. At first, range of number of downlink neighbors need to be fixed according to the current standard and it is as in equation (7).

$$ednb\_n_i = \begin{cases} ndnb\_n_i, & \text{if } ndnb\_n_i \leq 2ndnb\_CH \\ 2ndnb\_CH, & \text{otherwise} \end{cases}$$

Therefore, $\quad dnc = ndnb\_n_i / 2ndnb\_CH \quad (7)$

Where '*ednb_$n_i$*' is the effective downlink neighbors and '*dnc*' is the downlink neighbor connectivity, clearly range of '*dnc*' is between 0 to 1. It is also assumed that '*ndnb_CH*' is never be zero as before if become zero, the CH will be changed.

## 4. Several security issues in MANETs and respective our proposed solution

Besides other challenges such as dynamic topology, energy constrained, lack of bandwidth, MANETs faces two more serious challenges, which are selfish and malicious behavior of the node. These challenges may come from some node(s) within the network. According to behavior of the node, all nodes of the network can be classified into three categories, such as normal node, selfish node and malicious node. The normal node works in the expected way, therefore no need to be worry about it. Selfish node works in unexpected ways, whereas malicious node does more harmful for the network. This scheme proposed a solution to security threats come from malicious node(s) within the network. A malicious node gives main security challenges in MANETs. Different types of malicious attack and our proposed solution discussed below.

*4.1 Black Hole Attack*

Malicious node could participate in communication of other nodes by replying false route reply (RREP) packet with mentioning shortest path to the intended destination. The source node could fall on this trap of malicious node and start sending data packet through this malicious node. Therefore, malicious node will get chances to drop those packets or do some more harmful work such as tampering data packets etc.

In the figure-2 a portion of MANETs has shown, where malicious node is m, source node s and destination node is d. The source node s wants to communicate with destination node d. Therefore, node s broadcast route request (RREQ) packet to its neighbors including p, and p also broadcast this RREQ to its neighbor nodes same way. In this way malicious node m gets RREQ packet for destination node d, then node m could reply with a route reply (RREP) packet with false information by telling it has the shortest path to the destination d. The source node s believes that malicious node m, and start sending data packet towards m. Then m can drop those packets or do more harmful work on these data packets. Similarly reverse direction packet also can be captured by this malicious node.

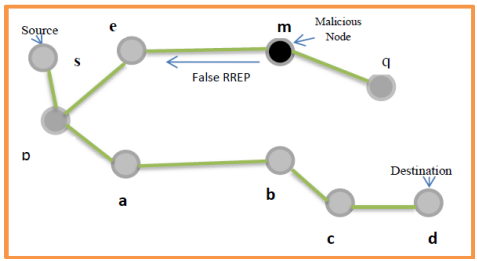 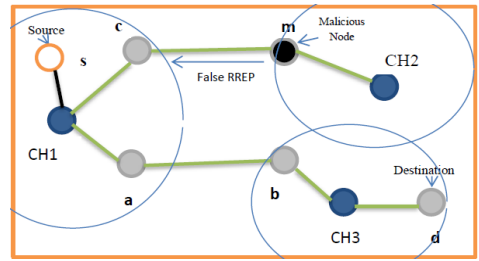

Figure-2: Black hole attack by node m    Figure-3: Solution to the black hole attack

This black hole attack can prevent through CH, or even arise, CH can give punishment to those malicious nodes and separate out from the network. As it discussed earlier in this clustering scheme, communication is performed through CH, Gateway node, source node and destination node. In this scheme as we mentioned, the black hole attack could arise by gateway node only, and between two neighbors CHs at most two gateway nodes could participate in a communication path. Gateway node is one-hop away from CH, Therefore, activity can be monitored by CH easily. If any of gateway nodes carry out a black hole attack, that is, its first reply a false information to enter in communication path, but if this node, starts dropping packet, then it can be easily caught on the red hand by consulting the neighbor CH. Then a malicious node can be separate out, not only from the cluster but from the network.

In the example mentioned in the figure-2, malicious node m can be bounded by two successive CHs, that are CH1 and CH2 which is shown in figure-3. Between these two CHs gateway nodes c and m can only assist communication between these two clusters headed by CH1 and CH2. Here node m is directly monitored by the cluster head CH1, therefore, node m has no chance to carry out the black hole attack as all communications controlled by only CHs in our scheme. Therefore, here this particular required communication can be done by CH1 and CH3.

*4.2 Wormhole attack:*

Here two successive malicious nodes collude and make a wormhole between them. Whenever a route request comes, colluding nodes hide their node information, so source node does not understand the presence of these two nodes. The source node estimates the path length, which are less than two hops to actual path length. Therefore, the probability of selection of this path is very high. If this path is selected, then source node start sending data packet through this path, then these two malicious nodes can drop these packets or do more harmful work such as tampering etc.

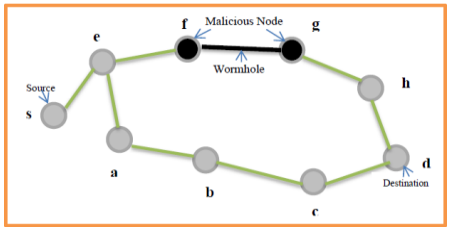     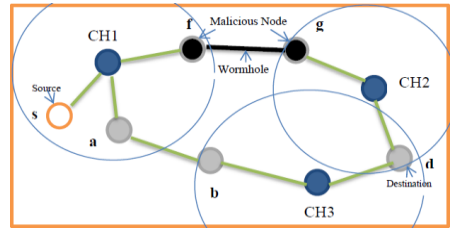

Figure-4: Wormhole attack by nodes f and g     Figure-5: Solution to the wormhole attack

In this figure-4 a portion of the networks shown, here source node s wants to communicate with destination node d. Source nodes broadcast RREQ packet to its neighbors, including e. Whenever we broadcast RREQ, node f receive it and it passes through the wormhole to nodes g, and nodes g sends it to node h without putting any marks. Therefore, he understands that this route request packet comes from node e directly, but it was not; in this way a virtual shortest path s-e-h-d established but actual path is s-e-f-g-h-d. Now source node s selects this virtual path s-e-h-d instead real shortest path s-a-b-c-d. Whenever a source node s starts sending data packets through this shortest path, malicious node able capture those data packets. Reverse direction data packet also can be captured by the same way.

As it is earlier mentioned that at most two gateway nodes could belongs in between two successive CHs. In wormhole attack two successive nodes collude to make a wormhole between them, therefore wormhole attack is not feasible in single hop clustering scheme. Because two gateway nodes directly monitoring by its CH, and that is the advantage of single hop clustering scheme. Even such kind of activity initiated by gateway nodes, it would easier to catch in red hand by two successive CHs. In the example mentioned in figure-4, malicious nodes f and g collude to make a wormhole between these two nodes. But if we see this portion of the networks through our scheme then look like in figure-5. Here node f directly controlled by the cluster head CH1 and g directly controlled by CH2. The nodes f and g cannot collude to make a wormhole without knowing their respective cluster heads. Therefore, wormhole attacks not feasible in our single hop clustering scheme.

### *4.3 Spoofing (Impersonation Attacks)*

Some malicious nodes hide their addresses and use address of another node during communication. After that malicious node do harm the network. Therefore, some normal node gets falsely blamed and loose trust level for such type of malicious node in the network. This type of attack is called spoofing.

In this scheme, every node just single hop away from its respective CH, and reply to HELLO messages time to time to respective CH. If any node tries to carry out spoofing attacks by hiding its address, then it CH will check the address or identity through the link the node use to connect its CH, and if CH found that the node is hiding it address, then CH will catch the malicious node on red handed from its cluster.

*4.4 Slander Attack:*

It is quite similar to spoofing, here malicious node attempt to reduce the overall trust of another node. But here malicious node doesn't hide its address, instead it collided with other malicious nodes. After that, start sending false information about a normal node to reduce their trust level within the networks.

A cluster member node might collude with other cluster member node and give false information to CH to reduce the trust value of the target node. But this is not possible for this scheme as explain in section-3; every node directly attached to respective CH; therefore, increase or decrease of trust value of a node directly done by respective CH without any certification of the other member node. So slander, attack can be resisted through CH.

*4.5 Routing Table Overflow Attack*

This type of attack occurs basically on proactive routing where routing table is maintained by each node even routes are not required. A malicious node sends false information by claiming it has many routes to many nodes, but actually those nodes are not exists. In this way malicious node try to overflow the routing tables of other nodes, so that later other node could not add more real route information into the routing table.

Some member node may unnecessarily send false information to its CH to overflow routing table of that CH, therefore, required routing information could drop by CH. But here CH only node in a cluster which maintain route and store routing table and as already mentioned that this scheme assume CH node are trust worthy. Therefore, no question arises of this kind attack in our routing scheme.

*4.6 Grey Hole Attack*

Here malicious node flow same principle of black hole attack, but here malicious node drops selective packet such as data packets, but let it go control packets such as RREQ packet. Therefore, another node falsely understands that malicious node positivity active in communication as normal node.

The Grey hole attack is very difficult to catch because here malicious node pass controlling packet as it mentioned. This scheme makes resistance to the grey hole as a black hole attack. Here also this can come only from gateway node. This scheme assures data packet delivery only after getting the acknowledgment message of respective data packet from neighbors CH. Therefore, this scheme can make resistances to grey hole attack.

## 5. Simulation

Simulation environment appears in table-1 for simulation experiments. Performance analysis of these algorithms is done using network simulation (NS-2) version 2.33. C3H is compared with CCS and EAACK which are two state-of-the art approached to detection of selfish and malicious nodes. Simulation metrics are a percentage of correct detection, malicious nodes, network throughput (percentage of data packets that could reach their respective destinations) and end-to-end delay per session.

In C3H, each CH computes the trust value of its members based on their previous activities and this trust value is considered along with residual energy and relative velocity. Therefore, if a node ceases to forward one particular message, the CH can easily investigate the chances of its complete exhaustion and breakage of links.

Table 1: Simulation Parameters

| Topology area | 500m × 500m |
|---|---|
| Traffic type | Constant bit rate (CBR) |
| Packet size | 512 bytes |
| HELLO packet interval for original versions of protocols | 10 ms |
| Node mobility | 10-30 m/s |
| Signal frequency | 2.4 GHz |
| Channel capacity | 2 Mbps |
| Transmission power | 300-600 mW |
| Receiving power | 50-300 mW |
| Mobility model | Random waypoint |
| Radio range | 50 – 100 m |
| Initial energy of nodes | 5 j – 10 j |
| Pause time | 1 s |
| Number of nodes | 20, 40, 60, 80, 100 |

Simulation graphs appear in figures 6, 7 and 8. Unlike CCS and EAACK, C3H considers residual energy of nodes and relative velocity between a CH and its members. If residual energy is very high and relativity is low, but still the node keeps mum to message forwarding request of its CH, then the node is accused of malicious activity and its trust value reduce. If this trust value reduces below a pre-defined limit, then the node is blacklisted network-wide. As seen in figure 6, correct detection of malicious activity is higher in case of C3H. Reason is that, its mentioned competitors do not consider factors like energy and velocity and therefore, sometimes punish non-malicious nodes which is not right. In that way, we lose links to certain good nodes and also packets generated by them are not forwarded to their respective destinations; no nodes cooperate with them. So, network throughput in C3H is much higher than CCS and EAACk as seen from figure 7. Figure 8 is concerned with end-to-end delay which is much less in C3H due to availability of a number of good links.

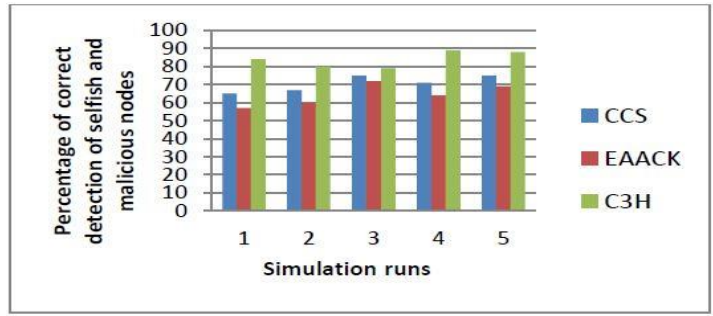

Figure 6: Percentage of correct detection of malicious activities in different simulation runs

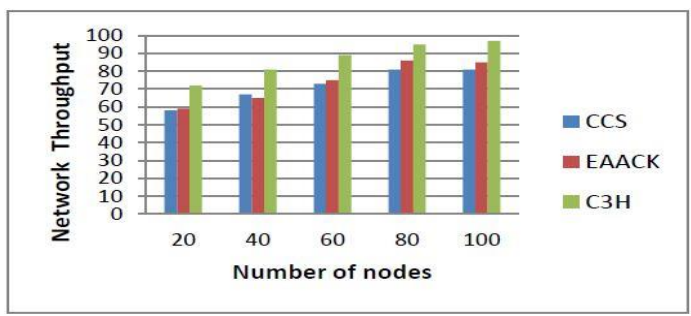

Figure 7: Network throughput vs number of nodes

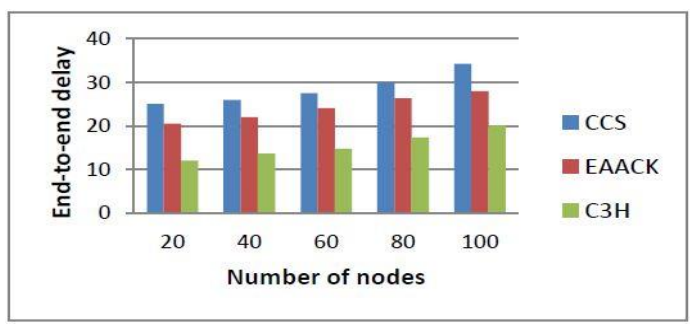

Figure 8: End-to-end delay per session vs number of nodes

## 6. Conclusion and future scope

In this scheme cluster head (CHs) are the most vital node as this scheme assumes CH is a most trustworthy node. Therefore, choosing the best candidate for cluster head is most important, and this scheme does the same using four parameters described in subsection 3.2. Then through these CHs malicious attacks can be avoided and prevented.  CH takes packet transfer requests and give processing priority based on trust values of that node. This clustering strategy could also be used to increase node co-operation, which is a very essential to successful data transmission in civilian data communication using MANETs.